\documentclass[aps,preprint]{revtex4}%
\usepackage{amsfonts}
\usepackage{amsmath}
\usepackage{amssymb}
\usepackage{graphicx}%
\newtheorem{theorem}{Theorem}
\newtheorem{acknowledgement}[theorem]{Acknowledgement}

\newtheorem{conjecture}[theorem]{Conjecture}

\newtheorem{proposition}[theorem]{Proposition}

\begin{document}
\accepted{Physical Review A}

\title[Macroscopic objects in quantum mechanics]{Macroscopic objects in quantum mechanics: A combinatorial approach}
\author{Itamar Pitowsky}
\affiliation{Department \ of Philosophy, The Hebrew University, Mount Scopus, Jerusalem
91905, Israel.}
\email{itamarp@vms.huji.ac.il}

\begin{abstract}
Why we do not see large macroscopic objects in entangled states? There are two
ways to approach this question. The first is dynamic: the coupling of a large
object to its environment causes any entanglement to decrease considerably.
The second approach, which is discussed in this paper, puts the stress on the
difficulty to \emph{observe} a large scale entanglement. As the number of
particles $n$ grows we need an ever more precise knowledge of the state, and
an ever more carefully designed experiment, in order to recognize
entanglement. To develop this point we consider the family of observables,
called witnesses, which are designed to detect entanglement. A witness $W$
distinguishes all the separable (unentangled) states from some entangled
states. If we normalize the witness $W$ to satisfy $\left\vert tr(W\rho
)\right\vert \leq1$ for all separable states $\rho$, then the efficiency of
$W$ depends on the size of its maximal eigenvalue in absolute value; that is,
its operator norm $\left\Vert W\right\Vert $. It is known that there are
witnesses on the space of $n$ qbits for which $\left\Vert W\right\Vert $ is
exponential in $n$. However, we conjecture that for a large majority of
$n$-qbit witnesses $\left\Vert W\right\Vert \leq O(\sqrt{n\log n})$. Thus, in
a non ideal measurement, which includes errors, the largest eigenvalue of a
typical witness lies below the threshold of detection. We prove this
conjecture for the family of extremal witnesses introduced by Werner and Wolf
\ (\emph{Phys. Rev. A }\textbf{64}, 032112 (2001)).

\end{abstract}
\maketitle

\section{Introduction}

Why we do not see large macroscopic objects in entangled states? There are two
ways to approach this question. The first is dynamic: the coupling of a large
object to its surroundings and its constant random bombardment from the
environment cause any entanglement that ever existed to disentangle. The
second approach- which does not in any way conflict with the first- puts the
stress on the word \emph{we}. Even if the particles composing the object were
all entangled and insulated from the environment, we shall still find it hard
to observe the superposition. The reason is that as the number of particles
$n$ grows we need an ever more precise knowledge of the state, and an ever
more carefully designed experiment in order to recognize the entangled
character of the state of the object.

In this paper I examine the second approach by considering entanglements of
multi-particle systems. For simplicity, the discussion concentrates mostly on
two-level systems, that is, $n$-qbit systems where $n$ is large. An observable
$W$ that distinguishes all the unentangled states from some entangled states
is called a \emph{witness}. For our purpose it is convenient to use the
following definition: A witness $W$ satisfies $\max\left\vert tr(W\rho
)\right\vert =1$, where the maximum is taken over all separable states $\rho$;
while $\left\Vert W\right\Vert >1$, where $\left\Vert W\right\Vert $ is the
operator norm, the maximum among the absolute values of the eigenvalues of
$W$. It is easy to see that this definition is equivalent to the usual one
\cite{1}.

For each $n\geq2$ there is a witness $W$, called the Mermin-Klyshko operator
\cite{2}, whose norm, $\left\Vert W\right\Vert =\sqrt{2^{n-1}}$, increases
exponentially with $n$. The eigenvector at which this norm obtains is called
the generalized GHZ state; it represents a maximally entangled system of $n$
qbits. However, we shall see that this large norm is the exception, not the
rule. My aim is to show that the norm of a \emph{majority }of the
witnesses\emph{\ }grows with $n$ very slowly, $\left\Vert W\right\Vert \leq
O(\sqrt{n\log n})$, slower than the growth of the measurement error. We shall
prove this with respect to a particular family of witnesses, and formulate it
as a conjecture in the general case. This means that unless the system has
been very carefully prepared in a specific state, there is a very little
chance that we shall detect multiparticle entanglement, even if it is there.

\section{Witnesses and Quantum Correlations}

\subsection{The two particles case}

Let $A_{0},A_{1},B_{0},B_{1}$ be four Hermitian operators in a finite
dimensional Hilbert space $\mathbb{H}$ such that $A_{i}^{2}=B_{j}%
^{2}=\mathbf{1}$. On $\mathbb{H\otimes H}$ define the following operator
\begin{equation}
W=\frac{1}{2}A_{0}\otimes B_{0}+\frac{1}{2}A_{0}\otimes B_{1}+\frac{1}{2}%
A_{1}\otimes B_{0}-\frac{1}{2}A_{1}\otimes B_{1} \label{1}%
\end{equation}
It is easy to see that $\left|  tr(W\rho)\right|  \leq1$ for any separable
state $\rho$ on $\mathbb{H\otimes H}$. This is, in fact, the Clauser Horne
Shimony and Holt (CHSH) inequality \cite{4}. Indeed, if $X_{0},X_{1}%
,Y_{0},Y_{1}$ are any four random variables taking the values $\pm1$ then
\begin{equation}
-1\leq\frac{1}{2}X_{0}Y_{0}+\frac{1}{2}X_{0}Y_{1}+\frac{1}{2}X_{1}Y_{0}%
-\frac{1}{2}X_{1}Y_{1}\leq1 \label{2}%
\end{equation}
as can easily be verified by considering the 16 possible cases. Hence
$c_{ij}=\mathcal{E(}X_{i}Y_{j})$, the correlations between $X_{i}$ and $Y_{j}%
$, also satisfy the inequalities%

\begin{equation}
-1\leq\frac{1}{2}c_{00}+\frac{1}{2}c_{01}+\frac{1}{2}c_{10}-\frac{1}{2}%
c_{11}\leq1 \label{3}%
\end{equation}
If $\rho$ is a separable state on $\mathbb{H\otimes H}$ we can represent
$tr(\rho$ $A_{i}\otimes B_{j})$ as correlations $c_{ij}$ between such $X_{i}%
$'s and $Y_{j}$'s. In other words, the correlations can be recovered in a
local hidden variables model.

From Eq~(\ref{2}) three other conditions of the form Eq~(\ref{3}) can be
obtained by permuting the two $X$ indices or the two $Y$ indices. The
resulting constraints on the correlations form a necessary and
\emph{sufficient} condition for the existence of a local hidden variables
model \cite{5,6}. One way to see this is to consider the four-dimensional
convex hull $C_{2}$ of the sixteen real vectors in $\mathbb{R}^{4}$:%

\begin{equation}
(X_{0}Y_{0},X_{0}Y_{1},X_{1}Y_{0},X_{1}Y_{1})\quad X_{i}=\pm1,\ Y_{j}=\pm1
\label{4}%
\end{equation}
and prove that an arbitrary four-dimensional real vector $(c_{00}%
,c_{01},c_{10},c_{11})$ is an element of $C_{2}$ if, and only if, the $c_{ij}$
satisfy $-1\leq c_{ij}\leq1$ and all the conditions of type Eq~(\ref{3}). The
inequalities, then, are the \emph{facets} of the polytope $C_{2} $.

By contrast with the classical case let $\rho$ be a \emph{pure}
\emph{entangled} state on $\mathbb{H\otimes H}$. Then for a suitable choice of
$A_{i}$'s, and $B_{j}$'s in Eq~(\ref{1}) we have $\left\vert tr(W\rho
)\right\vert >1$\cite{7}, so the operators $W$ of this type are sufficient as
witnesses for all pure entangled bipartite states. To obtain a geometric
representation of the quantum correlations let $\rho$ be any state, denote
$q_{ij}=tr(\rho$~$A_{i}\otimes B_{j})$, and consider the set $Q_{2}$ of all
four dimensional vectors $(q_{00},q_{01},q_{10},q_{11})$ which obtain as we
vary the Hilbert space and the choice of $\rho$, $A_{i}$, and $B_{j}$. The set
$Q_{2}$ is convex, $Q_{2}\supset C_{2}$ but it is not a polytope. The shape of
this set has been the focus of a great deal of interest \cite{8,9,10,11,12}.
To get a handle on its boundary we can check how the value of $\left\Vert
W\right\Vert $ changes with the choice of $A_{i}$'s, and $B_{j}$'s. Cirel'son
\cite{8,9} showed that $\left\Vert W\right\Vert \leq\sqrt{2}$. This is a tight
inequality, and equality obtains already for qbits. In this case,
$\mathbb{H=C}^{2}$ and $A_{i}=\sigma(\mathbf{a}_{i})$, $B_{j}=\sigma
(\mathbf{b}_{j})$ are spin operators, with $\mathbf{a}_{0}\mathbf{,a}%
_{1}\mathbf{,b}_{0}\mathbf{,b}_{1}$ four directions in physical space. There
is a choice of directions such that $\left\Vert W\right\Vert =\sqrt{2}$, and
the eigenvectors $\left\vert \phi\right\rangle \in\mathbb{C}^{2}%
\otimes\mathbb{C}^{2}$, corresponding to this value are the maximally
entangled states.

\subsection{The n particles case - Werner Wolf operators}

Some of these results can be extended to $n$ particle systems, provided that
the operators are restricted to two binary measurements per particle. To
account for classical correlations consider $2n$ random variables $X_{0}%
^{1},X_{1}^{1};X_{0}^{2},X_{1}^{2};...;X_{0}^{n},X_{1}^{n}$, each taking the
two possible values $\pm1$. We shall parametrize the coordinates of a vector
in the $2^{n}$-dimensional real space $\mathbb{R}^{2^{n}}$ by sequences
$\mathbf{s=}(s_{1},...,s_{n})\in\{0,1\}^{n}$. Now, consider the set of
$2^{2n}$ real vectors in $\mathbb{R}^{2^{n}}$%

\begin{equation}
\mathbf{(}a(0,...,0),...,a(s_{1},...,s_{n}),...,a(1,...,1)\mathbf{)},\quad
a(s_{1},...,s_{n})=X_{s_{1}}^{1}X_{s_{2}}^{2}...X_{s_{n}}^{n} \label{5}%
\end{equation}
Their convex hull in $\mathbb{R}^{2^{n}}$, denoted by $C_{n}$, is the range of
values of all possible classical correlations for $n$ particles and two
measurements per site. Werner and Wolf \cite{3}, and independently Zukowski
and Brukner \cite{13} showed that $C_{n}$ is a hyper-octahedron and derived
the inequalities of its facets. These are $2^{2^{n}}$ inequalities of the form%

\begin{equation}
-1\leq\sum_{s_{1},...,s_{n}=0,1}\beta_{f}(s_{1},...,s_{n})X_{s_{1}}%
^{1}X_{s_{2}}^{2}...X_{s_{n}}^{n}\leq1 \label{6}%
\end{equation}
where each inequality is determined by an arbitrary function $f:\{0,1\}^{n}%
\rightarrow\{-1,1\}$ with%

\begin{equation}
\beta_{f}(s_{1},...,s_{n})=\frac{1}{2^{n}}\sum_{\varepsilon_{1}%
,...,\varepsilon_{n}=0,1}(-1)^{\varepsilon_{1}s_{1}+...+\varepsilon_{n}s_{n}%
}f(\varepsilon_{1},...,\varepsilon_{n}) \label{7}%
\end{equation}
In other words, to each choice of function $f$ there corresponds a choice of
coefficients $\beta_{f}$. Since $\beta_{f}$ is the inverse Fourier transform
of $f$ on the group $\mathbb{Z}_{2}^{n}$ we have by Plancherel's theorem
\cite{14}:%

\begin{equation}
\sum_{\mathbf{s}}\left|  \beta_{f}(\mathbf{s})\right|  ^{2}=\frac{1}{2^{n}%
}\sum_{\mathbf{\varepsilon}}\left|  f(\mathbf{\varepsilon})\right|  ^{2}=1
\label{8}%
\end{equation}

Using the analogy with the bipartite case let $A_{0}^{1},A_{1}^{1}%
,...,A_{0}^{n},A_{1}^{n}$ be $2n$ arbitrary Hermitian operators in a Hilbert
space $\mathbb{H}$, satisfying $(A_{i}^{j})^{2}=\mathbf{1}$. The quantum
operators corresponding to the classical facets in Eq~(\ref{6}) are the
\emph{Werner Wolf operators} on $\mathbb{H}^{\otimes n}$\textit{\ }given by:%

\begin{equation}
W_{f}=\sum_{s_{1},...,s_{n}\in\{0,1\}}\beta_{f}(s_{1},...,s_{n})A_{s_{1}}%
^{1}\otimes...\otimes A_{s_{n}}^{n} \label{9}%
\end{equation}

It is easy to see from Eq~(\ref{6}) that $\left|  tr(\rho W_{f})\right|  \leq1
$ for every separable state $\rho$ and all the $f$'s. However, the
inequalities may be violated by entangled states. Let $Q_{n}$ be the set of
all vectors in $\mathbb{R}^{2^{n}}$ whose coordinates have the form
$q(s_{1},...,s_{n})=tr(\rho A_{s_{1}}^{1}\otimes...\otimes A_{s_{n}}^{n})$ for
some choice of state $\rho$ and operators $A_{i}^{j}$ as above. The set
$Q_{n}$ is the range of possible values of quantum correlations, and it is not
difficult to see that $Q_{n}$ is convex and $Q_{n}\supset C_{n}$. To obtain
information about the boundary of $Q_{n}$ we can examine how $\left\|
W_{f}\right\|  $ varies as we change the $A_{i}^{j}$'s. In this case too, it
was shown \cite{3} that for each fixed $f$, the maximal value of $\left\|
W_{f}\right\|  $ is already obtained when we choose $\mathbb{H=C}^{2}$, and
the $A_{i}^{j}$'s to be spin operators. Therefore, without loss of generality, consider%

\begin{equation}
W_{f}=\sum_{s_{1},...,s_{n}\in\{0,1\}}\beta_{f}(s_{1},...,s_{n})\sigma
(\mathbf{a}_{s_{1}}^{1})\otimes...\otimes\sigma(\mathbf{a}_{s_{n}}^{n})
\label{10}%
\end{equation}

Where $\mathbf{a}_{0}^{1},\mathbf{a}_{1}^{1},...,\mathbf{a}_{0}^{n}%
,\mathbf{a}_{1}^{n}$ are $2n$ arbitrary directions. We can calculate
explicitly the eigenvalues of $W_{f}$ \cite{3,15}. Let $\mathbf{z}_{j}$ be the
direction orthogonal to the vectors $\mathbf{a}_{0}^{j},\mathbf{a}_{1}^{j}$,
$j=1,...,n$. Denote by $\left\vert -1\right\rangle _{j}$ and $\left\vert
1\right\rangle _{j}$ the states \textquotedblleft spin-down\textquotedblright%
\ and \textquotedblleft spin-up\textquotedblright\ in the $\mathbf{z}_{j}%
$-direction; so the vectors $\left\vert \omega_{1},\omega_{2},...,\omega
_{n}\right\rangle $, $\omega=(\omega_{1},\omega_{2},...,\omega_{n}%
)\in\{-1,1\}^{n}$ form a basis for the $n$-qbits space. Let $\mathbf{x}_{j}$
be orthogonal to $\mathbf{z}_{j}$ and let $\theta_{s}^{j}$ be the angle
between $\mathbf{a}_{s}^{j}$ and $\mathbf{x}_{j}$, $s=0,1$. For each $f$ there
are $2^{n}$ eigenvectors of $W_{f}$ which have the generalized GHZ form%

\begin{equation}
\left|  \Psi_{f}(\omega)\right\rangle =\frac{1}{\sqrt{2}}(e^{i\Theta(\omega
)}\left|  \omega_{1},\omega_{2},...,\omega_{n}\right\rangle +\left|
-\omega_{1},-\omega_{2},...,-\omega_{n}\right\rangle ) \label{11}%
\end{equation}
and the corresponding eigenvalue
\begin{equation}
\lambda_{f}(\omega)=e^{i\Theta(\omega)}\sum_{s_{1},...,s_{n}\in\{0,1\}}%
\beta_{f}(s_{1},...,s_{n})\exp i\left(  \omega_{1}\theta_{s_{1}}%
^{1}+...+\omega_{n}\theta_{s_{n}}^{n}\right)  \label{12}%
\end{equation}
where $\Theta(\omega)$ in Eqs~(\ref{11},\ref{12}) is chosen so that
$\lambda_{f}(\omega)$ is a real number. Hence%

\begin{equation}
\left\|  W_{f}\right\|  =\underset{\omega}{\ \max}\left|  \lambda_{f}%
(\omega)\right|  \label{13}%
\end{equation}
As in the CHSH case we can check how large $\left\|  W_{f}\right\|  $ can
become as $\mathbf{a}_{0}^{j},\mathbf{a}_{1}^{j}$ range over all possible
directions. Using Eqs~(\ref{12},\ref{13}) we see that%

\begin{equation}
\underset{\mathbf{a}_{0}^{j},\mathbf{a}_{1}^{j}}{\max}\left\Vert
W_{f}\right\Vert =\underset{\theta_{0}^{1},\theta_{1}^{1},...,\ \theta_{0}%
^{n},\theta_{1}^{n}}{\max}\left\vert \sum_{s_{1},...,s_{n}\in\{0,1\}}\beta
_{f}(s_{1},...,s_{n})\exp i\left(  \theta_{s_{1}}^{1}+...+\theta_{s_{n}}%
^{n}\right)  \right\vert \label{14}%
\end{equation}
with the maximum on the left is taken over all possible choices of directions
$\mathbf{a}_{0}^{1},\mathbf{a}_{1}^{1},...,\mathbf{a}_{0}^{n},\mathbf{a}%
_{1}^{n}$. The Mermin-Klyshko operators \cite{2}, mentioned previously,
correspond to a particular choice of $f_{0}:\{0,1\}^{n}\rightarrow\{-1,1\}$
and $\mathbf{a}_{0}^{j},\mathbf{a}_{1}^{j}$, with the result that $\left\Vert
W_{f_{0}}\right\Vert =\sqrt{2^{n-1}}$. This is the maximal value of
Eq~(\ref{14}) possible. The maximum value is attained by a small minority of
the operators $W_{f}$, only those which are obtained from $W_{f_{0}}$ by one
of the $n!2^{2n+1}$ symmetry operations of the polytope $C_{n}$ (as compared
with the total of $2^{2^{n}}$ of facets in Eq~(\ref{6})).

In any case, for most $f$ 's there is a choice of angles such that $W_{f}$ is
a witness. This means that in addition to the fact that $\left\vert tr(\rho
W_{f})\right\vert \leq1$ for every separable state $\rho$, we also have
$\left\Vert W_{f}\right\Vert >1$. The $2^{n}$ exceptional cases are those in
which the inequality in Eq~(\ref{6}) degenerates into the trivial condition
$1\leq X_{s_{1}}^{1}X_{s_{2}}^{2}...X_{s_{n}}^{n}\leq1$. All the other $W_{f}%
$'s are witnesses. The reason is that all inequalities of type Eq~(\ref{6})
are obtained from the basic inequalities for $C_{2}$ (including the trivial
ones) by iteration \cite{3}. If the iteration contains even one instance of
the type Eq~(\ref{2}) the corresponding operator can be chosen to violate the
CHSH inequality.

\section{Random witnesses}

\subsection{Typical behavior of $\ \max_{\mathbf{a}_{0}^{j},\mathbf{a}_{1}%
^{j}}\left\Vert W_{f}\right\Vert $}

Although the norm of $\left\Vert W_{f}\right\Vert $ can reach as high as
$\sqrt{2^{n-1}}$ this is not the rule but the exception. Our aim is to
estimate the \emph{typical} behavior of $\max_{\mathbf{a}_{0}^{j}%
,\mathbf{a}_{1}^{j}}\,\left\Vert W_{f}\right\Vert $ as we let $f$ range over
all its values (and the maximum taken over all directions $\mathbf{a}_{0}%
^{j},\mathbf{a}_{1}^{j}$). To do that, consider the set of all $2^{2^{n}}$
functions $f:\{0,1\}^{n}\rightarrow\{-1,1\}$ as a probability space with a
uniform probability distribution $\mathcal{P}$, which assigns probability
$2^{-2^{n}}$ to each one of the $f$'s. Then, for each set of fixed directions
$\mathbf{a}_{i}^{j}$ we can look at $\left\Vert W_{f}\right\Vert $ as a random
variable defined on the space of $f$'s. Likewise, also $\max_{\mathbf{a}%
_{0}^{j},\mathbf{a}_{1}^{j}}\,\left\Vert W_{f}\right\Vert $ in Eq~(\ref{14})
is a random variable on the space of $f$'s, for which we have

\begin{theorem}
There is a universal constant $C$ such that
\begin{equation}
\mathcal{P\ }\{f;\ \max_{\mathbf{a}_{0}^{j},\mathbf{a}_{1}^{j}}\,\left\Vert
W_{f}\right\Vert >C\sqrt{n\log n}\ \}\rightarrow0\quad as\ n\rightarrow
\infty\label{15}%
\end{equation}

\end{theorem}

The proof of this result is based on the theorem of Salem, Zygmund, and Kahane
\cite{16} and is given in the appendix. This means that for the vast majority
of the $f$'s the violation of the classical inequalities Eq~(\ref{6}) is
small. Accordingly, the boundary of $Q_{n}$ is highly uneven about the facets
of $C_{n}$; it does not extend far above most of the facets of $C_{n}$, but
occasionally it has an extended exponential hump.

The expected growth of $\left\vert \lambda_{f}\right\vert $ is even slower
when the directions $\mathbf{a}_{i}^{j}$ (or the angles $\theta_{i}^{j}$) are
fixed. As a direct consequence of Tchebychev's inequality \cite{17} we get:

\begin{proposition}
For $\lambda_{f}$ in Eq~(\ref{12}) we have for all $M>1$ : $\mathcal{P\ }%
\{f;\ \left|  \lambda_{f}\right|  >M\}\leq\frac{1}{M^{2}}$.
\end{proposition}

(See the appendix for details). This means that most of the eigenvalues of the
$W_{f}$'s are bounded within a small sphere. The application of a randomly
chosen $W_{f}$ to any of its eigenstates $\left\vert \Psi_{f}(\omega
)\right\rangle $ in Eq~(\ref{12}) is unlikely to reveal a significant
violation of Eq~(\ref{6}).

\subsection{The random witness conjecture}

For appropriate choices of angles the Werner-Wolf operators Eq~(\ref{10}) are,
with very few exceptions, entanglement witnesses on the space of $n$ qbits.
They are very special witnesses for two reasons: firstly, they are local
operators. This means that if we posses many copies of a system made of
$n$-qbits, all in the same state $\left\vert \Phi\right\rangle $, we can
measure the expectation $\left\langle \Phi\left\vert W_{f}\right\vert
\Phi\right\rangle $ by performing separate measurements on each qbit of the
system. Secondly, even as local observables the Werner Wolf operators are
special, because of the restriction to two measurements per particle. Indeed,
one would have liked to extend the results beyond this restriction, and obtain
all the inequalities for any number of measurement per site, but this problem
is $NP$-hard even for $n=2$, see \cite{18}.

However, the $W_{f}$'s are the most likely to be violated among the local
operators with two measurements per site, because they are derived from the
facets of $C_{n}$. Moreover, we already noted that all the norm estimates are
also valid for the wider family given in Eq~(\ref{9}), with the $A_{j}^{i}$'s
acting on any finite dimensional space, and satisfying $(A_{j}^{i})^{2}=I$.
Hence, the estimate of theorem 1 includes many more witnesses than those given
in Eq~(\ref{10}). To an $n$ qbits system we can add auxiliary particles and
use quantum and classical communication protocols. As long as our overall
measurement is in the closed convex hull of operators of the form%

\begin{equation}
W=\sum_{s_{1},...,s_{k}\in\{0,1\}}\beta_{f}(s_{1},...,s_{k})A_{s_{1}}%
^{1}\otimes...\otimes A_{s_{k}}^{k} \label{16}%
\end{equation}
with the $A_{j}^{i}$'s satisfying $(A_{j}^{i})^{2}=I$, and $k\leq O(n)$, the
estimate of theorem 1 holds.

Hence, there is a reason to suspect that the typical behavior of the Werner
Wolf operators is also typical of general random witnesses. A random witness
is an observable drawn from the set of all witnesses $\mathcal{W}$ with
uniform probability. It is easy to give an abstract description of
$\mathcal{W}$: Consider the space of Hermitian operators on $(\mathbb{C}%
^{2})^{\otimes n}$, it has dimension $d_{n}=2^{n-1}(2^{n}+1)$. For an
Hermitian operator $A$ define the norm $\left[  \left[  A\right]  \right]
=\sup\left\|  A\left|  \alpha_{1}\right\rangle ...\left|  \alpha
_{n}\right\rangle \right\|  $ where the supremum ranges over all choices of
unit vectors\emph{\ }$\left|  \alpha_{i}\right\rangle \in\mathbb{C}^{2}$.
Denote the unit sphere in this norm by $\mathcal{K}_{1}=\{A;\ \left[  \left[
A\right]  \right]  =1\}$. It is a hypersurface of dimension $d_{n}-1$, which
is equipped with the uniform (Lebesgue) measure, and its total hyper-area is
finite. Now, denote by $\mathcal{K}_{2}$ the normal unit sphere $\mathcal{K}%
_{2}=\{A;\ \left\|  A\right\|  =1\}$. Here, as usual, $\left\|  A\right\|
=\sup\left\|  A\left|  \Phi\right\rangle \right\|  $ is the operator norm,
where the supremum is taken over all unit vectors $\left|  \Phi\right\rangle
$. The set of witnesses is $\mathcal{W=K}_{1}\setminus\mathcal{K}_{2}$. Note
that if $A\in\mathcal{W}$ then necessarily $\left\|  A\right\|  >1$. It is not
difficult to see that $\mathcal{W} $ is relatively open in $\mathcal{K}_{1}$,
and therefore has a non zero measure in $\mathcal{K}_{1}$. We consider the set
of all witnesses $\mathcal{W}$ on $(\mathbb{C}^{2})^{\otimes n}$ and the
normalized Lebesgue measure $\mathcal{P}$ on it.

\begin{conjecture}
There is a universal constant $C$ such that
\begin{equation}
\mathcal{P\ }\{W\in\mathcal{W};\quad\left\Vert W\right\Vert >C\sqrt{n\log
n}\ \}\rightarrow0\quad as\ n\rightarrow\infty. \label{17}%
\end{equation}

\end{conjecture}

Let $\left\vert \Phi\right\rangle $ be a unit vector in $(\mathbb{C}%
^{2})^{\otimes n}$. Consider the measure of entanglement defined by:%
\begin{equation}
\mathcal{E}(\left\vert \Phi\right\rangle )=\sup\{\left\Vert W\left\vert
\Phi\right\rangle \right\Vert ;\;W\in\mathcal{W}\}. \label{18}%
\end{equation}
Hence, $\mathcal{E}(\left\vert \Phi\right\rangle )$ is the least upper bound
on the expectation values that any witness can have in the state $\left\vert
\Phi\right\rangle $. Because we have normalized the witnesses by $\left[
\left[  W\right]  \right]  =1$ the measure $\mathcal{E}(\left\vert
\Phi\right\rangle )$ is finite for all $\left\vert \Phi\right\rangle $. The
"dual" of Eq~(17) is

\begin{conjecture}
There is a universal constant $C$ such that%
\begin{equation}
\mathcal{M\ }\{\left\vert \Phi\right\rangle ;\quad\mathcal{E}(\left\vert
\Phi\right\rangle )>C\sqrt{n\log n}\ \}\rightarrow0\quad as\ n\rightarrow
\infty\label{19}%
\end{equation}
Where $\mathcal{M}$ is the normalized Lebesgue measure on the unit sphere of
$(\mathbb{C}^{2})^{\otimes n}$, and the supremum is taken over all unit
vectors $\left\vert \Phi\right\rangle \in(\mathbb{C}^{2})^{\otimes n}$
\end{conjecture}

\emph{Conjecture 3} is based on the assumption that the witnesses of the type
$W_{f}$ form a sufficiently dense mesh in the $\left[  \left[  W\right]
\right]  $-norm on $\mathcal{W}$. If this is the case the conjectures may be
proved as a concentration effect of the kind expressed by Levy's lemma. This
lemma has been recently used in the theory of quantum information \cite{19,
20}.

\subsection{Discussion}

Assume that \emph{conjecture 3} is valid and consider the following highly
ideal situation: A macroscopic object (a single copy of it) is prepared in a
state unknown to us and is carefully kept insulated from environmental
decoherence. Since the state is unknown we choose randomly a witness $W$ to
examine it. Now, suppose that by pure luck the system happens to be in an
eigenstate of $W$; in fact, the eigenstate corresponding to its maximum
eigenvalue (in absolute value). This means, in particular, that the state of
the system is entangled; but can we detect this fact using $W$? A measurement
of a witness \emph{on a single copy} is a very complicated affair (just think
about any one of the $W_{f}$'s in Eq~(\ref{10})). Such a measurement
invariably involves manipulations of the individual particles. If we make the
reasonable assumption that each such manipulation introduces a small
independent error, we obtain a total measurement error that grows
exponentially with $n$. By the random witness conjecture, Eq~(\ref{17}), this
means that we are unlikely to see a clear non-classical effect. The typical
witness is a poor witness.

A natural question to ask is why should we consider the uniform measure over
witnesses as the correct probability measure. In other words, why is a witness
chosen at random according to the uniform measure typical? The answer is
implicit in the situation just described; we assume that we lack any knowledge
of the state, and therefore have no reason to give a preference to one witness
over another. The point is that our choice is not going to work \emph{even if
we were lucky.} At any rate,\emph{ }cases in which very little is known about
the quantum state of a macroscopic object are not rare.

Moreover, there is a reason to believe that a result about the uniform
distribution is also relevant to the case where partial information about the
state is available. In this case we should modify the uniform distribution and
condition it on the additional available constraints. However, there is an
exponential gap between the typical witness norm in Eq~(\ref{17}) and the norm
required for a successful measurement. This means that the additional
information should be pretty accurate to be of any help. Moreover, it is
possible that we will not be able to witness the entanglement of many
\emph{precisely known} states, because, quite likely, even the best witnesses
for such states have small norms.

\emph{Conjecture 4}, if true, is even more relevant to our ability to observe
macroscopic entanglement. There are two types of macroscopic or mesoscopic
states whose entanglement might be witnessed, and the conjecture concerns the
second case:

\textbf{1}. There may be relatively rare cases in which the entanglement
witness happens to be a thermodynamic observable, that is, an observable whose
measurement does not require manipulation of individual particles but only the
observation of some global property of the system. There are some indications
that this may be the case for some spin chains and lattices \cite{21}.

\textbf{2}. Cases of very strong entanglement, like the GHZ state in
Eq~(\ref{11}), which do require many manipulations of individual particles to
witness the entanglement of a single copy of it; however, the value of
$\mathcal{E}(\left\vert \Phi\right\rangle )$ is large enough to give
significant results that rise above the measurement errors.

This means that the answer to the question "why don't I see cats in
superposition" is twofold: decoherence surely, but even if we could turn it
off, there is the combinatorial possibility that "seeing" something like this
is nearly impossible. All this, luckily, does not prevent the existence of
exotic macroscopic superpositions that can be recorded.

There is some analogy between the present approach to multiparticle systems
and the point made by Khinchin on the foundations of classical statistical
mechanics \cite{22, 23}. While thermodynamic equilibrium has its origins in
the dynamics of the molecules, much of the \emph{observable} qualities of
multiparticle systems can be explained on the basis of the law of large
numbers. The tradition which began with Boltzmann identifies equilibrium with
ergodicity. The condition of ergodicity ensures that every integrable function
has identical phase-space and long-time averages. However, Khinchine points
out that this is an overkill, because most of the integrable functions do not
correspond to macroscopic (that is, thermodynamic) observables. If we
concentrate on thermodynamic observables, which involve averages over an
enormous number of particles, weaker dynamical assumptions will do the job.

I believe that a similar answer can be given to our original question, namely,
why we do not normally see large macroscopic objects in entangled states.
Since decoherence cannot be \textquotedblleft turned off\textquotedblright%
\ the multiparticle systems that we encounter are never maximally entangled.
But even if the amount of entanglement that remains in them is still
significant, \emph{we} cannot detect it, because the witnesses are simply too
weak and their states not sufficiently entangled.

\section{Appendix}

In the proof of theorem 1 we shall rely on a theorem in Fourier analysis due
to Salem, Zygmund and Kahane \cite{16}. Our aim is to consider random
trigonometric polynomials. So let $(\Omega,\Sigma,\mathcal{P})$ be a
probability space, where $\Omega$ is a set, $\Sigma$ a $\sigma$-algebra of
subsets of $\Omega$, and $\mathcal{P}:\Sigma\rightarrow\lbrack0,1]$ a
probability measure. For a random variable $\xi$ on $\Omega$ denote by
$\mathcal{E}(\xi)=\int_{\Omega}\xi(\omega)d\mathcal{P}(\omega)$ the
expectation of $\xi$. A real random variable $\xi$ is called \emph{subnormal
}if $\mathcal{E}(\exp(\lambda\xi))\leq\exp(\frac{\lambda^{2}}{2})$ for all
$-\infty<\lambda<\infty$.

A trigonometric polynomial in $r$ variables is a function on the torus
$\mathbb{T}^{r}$ given by
\begin{equation}
g(\mathbf{t)=}g(t_{1},t_{2},...,t_{r})=\sum b(k_{1},k_{2},...,k_{r}%
)e^{i(k_{1}t_{1}+k_{2}t_{2}+...+k_{r}t_{r})} \label{20}%
\end{equation}
where the sum is taken over all negative and nonnegative integers $k_{1}%
,k_{2},...,k_{r}$ which satisfy $\left\vert k_{1}\right\vert +\left\vert
k_{2}\right\vert +...+\left\vert k_{r}\right\vert \leq N$. The integer $N$ is
called \emph{the degree of the polynomial}. Denote $\left\Vert g\right\Vert
_{\infty}=\underset{t_{1},...,t_{r}}{\max}\left\vert g(t_{1},t_{2}%
,...,t_{r})\right\vert $.

\begin{theorem}
(Salem, Zygmund, Kahane) Let the $\xi_{j}(\omega)$, $j=1,2,...,J$ be a finite
sequence of real, independent, subnormal random variables on $\Omega$. Let
$g_{j}(\mathbf{t})$, $j=1,2,...,J$ be a sequence of trigonometric polynomials
in $r$ variables whose degree is less or equal $N$, and such that $\sum
_{j}\left\vert g_{j}(\mathbf{t})\right\vert ^{2}\leq1$ for all $\mathbf{t}$.
Then
\begin{equation}
\mathcal{P\ }\left\{  \omega;\quad\left\Vert \sum_{j=1}^{J}\xi_{j}%
(\omega)g_{j}(\mathbf{t})\right\Vert _{\infty}>C\sqrt{r\log N}\right\}
\leq\frac{1}{N^{2}e^{r}} \label{21}%
\end{equation}
for some universal constant $C$.
\end{theorem}

Note that the formulation here is slightly different from that in \cite{16},
but the proof is identical. Our probability space $\Omega$ is the set of all
functions $f:\{0,1\}^{n}\rightarrow\{-1,1\}$ with the uniform distribution
which assigns each such function $f$ a weight $2^{-2^{n}}$. On this space
consider the $2^{n}$ random variables $\xi_{\varepsilon}(f)$ defined for each
$\varepsilon=(\varepsilon_{1},...,\varepsilon_{n})\in\{0,1\}^{n}$ by
\begin{equation}
\xi_{\varepsilon}(f)=f(\mathbf{\varepsilon}) \label{23}%
\end{equation}
Now, note that $f\in\Omega$ iff $-f\in\Omega$ hence for each fixed
$\mathbf{\varepsilon}$ we get $\mathcal{E}(\xi_{\varepsilon})=2^{-2^{n}}%
\sum_{f}f(\mathbf{\varepsilon})=-2^{-2^{n}}\sum_{f}f(\mathbf{\varepsilon
})=-\mathcal{E}(\xi_{\varepsilon})$, and therefore $\mathcal{E}(\xi
_{\varepsilon})=0$, similarly, for $\mathbf{\varepsilon,\varepsilon}^{\prime}$
we have $\mathcal{E}(\xi_{\varepsilon}\xi_{\varepsilon^{\prime}}%
)=\delta(\mathbf{\varepsilon},\mathbf{\varepsilon}^{\prime})$ and so on; the
$2^{n}$ random variables $\xi_{\varepsilon}(f)$ are independent. Now, by a
similar argument%

\begin{align}
&  \mathcal{E}(\exp(\lambda\xi_{\varepsilon}))=2^{-2^{n}}\sum_{f}\exp(\lambda
f(\mathbf{\varepsilon}))=\smallskip\label{24}\\
&  2^{-2^{n}}\sum_{f}\frac{1}{2}[\exp(\lambda f(\mathbf{\varepsilon}%
))+\exp(-\lambda f(\mathbf{\varepsilon}))]=\frac{1}{2}(e^{\lambda}%
+e^{-\lambda})\leq e^{\frac{\lambda^{2}}{2}}\nonumber
\end{align}

To define the trigonometric polynomials note that by Eqs~(\ref{7},\ref{12})%

\begin{align}
&  \sum_{s_{1},...,s_{n}\in\{0,1\}}\beta_{f}(s_{1},...,s_{n})\exp i\left(
t_{s_{1}}^{1}+...+t_{s_{n}}^{n}\right)  =\label{25}\\
&  =\frac{1}{2^{n}}\sum_{\mathbf{\varepsilon}}f(\mathbf{\varepsilon}%
)\sum_{\mathbf{s}}(-1)^{^{\varepsilon_{1}s_{1}+...+\varepsilon_{n}s_{n}}}\exp
i\left(  t_{s_{1}}^{1}+t_{s_{2}}^{2}+...+t_{s_{n}}^{n}\right)  =\nonumber\\
&  =\sum_{\mathbf{\varepsilon}}f(\mathbf{\varepsilon})\frac{1}{2^{n}}%
\prod\limits_{j=1}^{n}\left(  \exp it_{0}^{j}+(-1)^{\varepsilon_{j}}\exp
it_{1}^{j}\right)  =\sum_{\mathbf{\varepsilon}}\xi_{\mathbf{\varepsilon}%
}(f)g_{\mathbf{\varepsilon}}(\mathbf{t})\nonumber
\end{align}
with
\begin{equation}
g_{\varepsilon}(\mathbf{t})=2^{-n}\prod_{j}\left(  \exp it_{0}^{j}%
+(-1)^{\varepsilon_{j}}\exp it_{1}^{j}\right)  \label{26}%
\end{equation}
The polynomials $g_{\varepsilon}(\mathbf{t})$ do not depend on $f$, have $2n$
variables $t_{0}^{j}$, $t_{1}^{j}$, $j=1,2,...,n$, and their degree is $n$. We
shall prove that $\sum_{\mathbf{\varepsilon}}\left\vert g_{\varepsilon
}(\mathbf{t})\right\vert ^{2}=1$ for all $\mathbf{t}$. Indeed, $\left\vert
g_{\varepsilon}(\mathbf{t})\right\vert ^{2}=2^{-2n}\left\vert \prod_{j}\left(
1+(-1)^{\varepsilon_{j}}\exp(i\phi_{j}\right)  )\right\vert ^{2}$, with
$\phi_{j}=t_{1}^{j}-t_{0}^{j}$. But, $\left\vert 1+\exp i\phi_{j}\right\vert
^{2}=4\cos^{2}\left(  \frac{\phi_{j}}{2}\right)  $ and $\left\vert 1-\exp
i\phi_{j}\right\vert ^{2}=4\sin^{2}\left(  \frac{\phi_{j}}{2}\right)  $ and
therefore $\sum_{\mathbf{\varepsilon}}\left\vert g_{\varepsilon}%
(\mathbf{t})\right\vert ^{2}=\prod_{j}\left(  \cos^{2}\left(  \frac{\phi_{j}%
}{2}\right)  +\sin^{2}\left(  \frac{\phi_{j}}{2}\right)  \right)  =1$.

From Eqs~(\ref{14},\ref{25}) we get%

\begin{equation}
\underset{\mathbf{a}_{0}^{j},\mathbf{a}_{1}^{j}}{\max}\left\Vert
W_{f}\right\Vert =\left\Vert \sum_{\mathbf{\varepsilon}}\xi
_{\mathbf{\varepsilon}}(f)g_{\mathbf{\varepsilon}}(\mathbf{t})\right\Vert
_{\infty} \label{27}%
\end{equation}

Hence, we can apply the Salem Zygmund Kahane inequality Eq~(\ref{19}) to the
present case, with $N=n$ and $r=2n$, to obtain theorem 1.

To prove proposition 2 consider $\left\vert \lambda_{f}\right\vert $ as a
random variable on the space of $f$'s. By Eq~(\ref{12}) we get
\begin{equation}
\left\vert \lambda_{f}\right\vert =\left\vert \sum_{s_{1},...,s_{n}\in
\{0,1\}}\beta_{f}(s_{1},...,s_{n})\exp i\left(  t_{s_{1}}^{1}+...+t_{s_{n}%
}^{n}\right)  \right\vert \label{28}%
\end{equation}
with $t_{s_{j}}^{j}=\omega_{j}\theta_{s_{j}}^{j}$. By Eq~(\ref{22}) we get
$\left\vert \lambda_{f}\right\vert =\left\vert \sum_{\mathbf{\varepsilon}}%
\xi_{\mathbf{\varepsilon}}(f)g_{\mathbf{\varepsilon}}(\mathbf{t})\right\vert
$. But $\mathcal{E}(\sum_{\mathbf{\varepsilon}}\xi_{\mathbf{\varepsilon}%
}g_{\mathbf{\varepsilon}})=0$, and $\mathcal{E}(\left\vert \sum
_{\mathbf{\varepsilon}}\xi_{\mathbf{\varepsilon}}g_{\mathbf{\varepsilon}%
}\right\vert ^{2})=\sum_{\mathbf{\varepsilon}}\left\vert g_{\varepsilon
}\right\vert ^{2}=1$. Therefore, by Tchebishev's inequality \cite{17} we have
$\mathcal{P\ }\{f;\ \left\vert \lambda_{f}\right\vert >M\}\leq\frac{1}{M^{2}}$
for all $M>1$.

\begin{acknowledgement}
The research for this paper was done while I was visiting the Perimeter
Institute, I am grateful for the hospitality. This research is supported by
the Israel Science Foundation, grant number 879/02.
\end{acknowledgement}

\end{document}